\documentclass[10pt,twoside,twocolumn]{IEEEtran}
\newcommand\nc\newcommand
\usepackage{times,amssymb,amsmath,epsfig,nicefrac,euscript,cite,mathrsfs,
balance,float,bbm}

\newtheorem{theorem}{Theorem}[section]

\newtheorem{lemma}[theorem]{Lemma}
\newtheorem{proposition}[theorem]{Proposition}

\newtheorem{example}{Example}
\newcommand{\ol}{\setlength{\itemsep}{-40pt}\begin{enumerate}}
\newcommand{\eol}{\end{enumerate}\setlength{\itemsep}{-\parsep}}

\nc\bfa{{\boldsymbol a}}\nc\bfA{{\bf A}}\nc\cA{{\mathcal A}}\nc\sA{{\mathscr A}}
\nc\bfb{{\boldsymbol b}}\nc\bfB{{\bf B}}\nc\cB{{\mathcal B}}\nc\sB{{\mathscr B}}
\nc\bfc{{\boldsymbol c}}\nc\bfC{{\bf C}}\nc\cC{{\mathcal C}}\nc\sC{{\mathscr C}}
\nc\bfd{{\boldsymbol d}}\nc\bfD{{\bf D}}\nc\cD{{\mathcal D}}\nc\sD{{\mathscr D}}
\nc\bfe{{\boldsymbol e}}\nc\bfE{{\bf E}}\nc\cE{{\mathcal E}}\nc\sE{{\mathscr E}}
\nc\bff{{\boldsymbol f}}\nc\bfF{{\bf F}}\nc\cF{{\mathcal F}}\nc\sF{{\mathscr F}}
\nc\bfg{{\boldsymbol g}}\nc\bfG{{\bf G}}\nc\cG{{\mathcal G}}\nc\sG{{\mathscr G}}
\nc\bfh{{\boldsymbol h}}\nc\bfH{{\bf H}}\nc\cH{{\mathcal H}}\nc\sH{{\mathscr H}}
\nc\bfi{{\boldsymbol i}}\nc\bfI{{\bf I}}\nc\cI{{\mathcal I}}\nc\sI{{\mathscr I}}
\nc\bfj{{\boldsymbol j}}\nc\bfJ{{\bf J}}\nc\cJ{{\mathcal J}}\nc\sJ{{\mathscr J}}
\nc\bfk{{\boldsymbol k}}\nc\bfK{{\bf K}}\nc\cK{{\mathcal K}}\nc\sK{{\mathscr K}}
\nc\bfl{{\boldsymbol l}}\nc\bfL{{\bf L}}\nc\cL{{\mathcal L}}\nc\sL{{\mathscr L}}
\nc\bfm{{\boldsymbol m}}\nc\bfM{{\bf M}}\nc\cM{{\mathcal M}}\nc\sM{{\mathscr M}}
\nc\bfn{{\boldsymbol n}}\nc\bfN{{\bf N}}\nc\cN{{\mathcal N}}\nc\sN{{\mathscr N}}
\nc\bfo{{\boldsymbol o}}\nc\bfO{{\bf O}}\nc\cO{{\mathcal O}}\nc\sO{{\mathscr O}}
\nc\bfp{{\boldsymbol p}}\nc\bfP{{\bf P}}\nc\cP{{\mathcal P}}\nc\sP{{\mathscr P}}
\nc\bfq{{\boldsymbol q}}\nc\bfQ{{\bf Q}}\nc\cQ{{\mathcal Q}}\nc\sQ{{\mathscr Q}}
\nc\bfr{{\boldsymbol r}}\nc\bfR{{\bf R}}\nc\cR{{\mathcal R}}\nc\sR{{\mathscr R}}
\nc\bfs{{\boldsymbol s}}\nc\bfS{{\bf S}}\nc\cS{{\mathcal S}}\nc\sS{{\mathscr S}}
\nc\bft{{\boldsymbol t}}\nc\bfT{{\bf T}}\nc\cT{{\mathcal T}}\nc\sT{{\mathscr T}}
\nc\bfu{{\boldsymbol u}}\nc\bfU{{\bf U}}\nc\cU{{\mathcal U}}\nc\sU{{\mathscr U}}
\nc\bfv{{\boldsymbol v}}\nc\bfV{{\bf V}}\nc\cV{{\mathcal V}}\nc\sV{{\mathscr V}}
\nc\bfw{{\boldsymbol w}}\nc\bfW{{\bf W}}\nc\cW{{\mathcal W}}\nc\sW{{\mathscr W}}
\nc\bfx{{\boldsymbol x}}\nc\bfX{{\bf Z}}\nc\cX{{\mathcal X}}\nc\sX{{\mathscr X}}
\nc\bfy{{\boldsymbol y}}\nc\bfY{{\bf Y}}\nc\cY{{\mathcal Y}}\nc\sY{{\mathscr Y}}
\nc\bfz{{\boldsymbol z}}\nc\bfZ{{\bf Z}}\nc\cZ{{\mathcal Z}}\nc\sZ{{\mathscr Z}}

\newcommand\reals{{\mathbb R}}

\newcommand\integers{{\mathbb Z}}

\nc\uz{\underline z}

\nc\qed{\mbox{\rule[0pt]{0.5ex}{1.3ex}}}
\newcommand{\remove}[1]{}

\begin{document}

\title{On the Number of Errors Correctable with Codes on Graphs
\thanks{The results of this paper were presented in part at the
2009 IEEE International Symposium on Information Theory, Seoul, Korea, July 2009.}}

\author{Alexander~Barg and Arya Mazumdar
\thanks{Alexander Barg is with the Department of Electrical
    and Computer Engineering and Institute for Systems Research,
    University of Maryland, College Park, MD 20742 and Institute for
    Problems of Information Transmission, Moscow, Russia (e-mail:
    abarg@umd.edu).}  \thanks{Arya Mazumdar is with the Department of
    Electrical and Computer Engineering and Institute for Systems
    Research, University of Maryland, College Park, MD 20742 (e-mail:
    arya@umd.edu).}  \thanks{Research supported in part by NSF grants
    CCF0830699, CCF0635271, DMS0807411, CCF0916919.} }

\maketitle

\begin{abstract}
We study ensembles of codes on graphs (generalized low-density parity-check, or
LDPC codes) 
constructed from random graphs and fixed local constrained codes,
and their extension to codes on hypergraphs.
It is known that the average minimum
distance of codes in these ensembles grows linearly with the code
length. We show that these codes can correct a linearly growing number
of errors under simple iterative decoding algorithms. In particular,
we show that this property extends to codes constructed by parallel
concatenation of Hamming codes and other codes with small minimum
distance. Previously known results that proved this property for
graph codes relied on graph expansion
and required the choice of local codes with large distance relative to
their length.
\end{abstract}


\section{Introduction}
Considerable attention in recent years was devoted to the study of error
correction with codes on graphs. In this paper we are interested in
estimating the number of errors correctable with codes on
graphs constructed as generalizations of LDPC codes. LDPC
codes are constructed on a bipartite graph $G(V,E), V=V_1\cup V_2$
by associating
code's coordinates with the vertices in one part of $G$,
replicating the values of each vertex on the edges incident to it,
and imposing a parity-check constraint at each vertex of the
other part of $G$.
The generalization that we have in mind is concerned with replacing
the repetition and single-parity-check codes as local codes at the
graph's vertices with other error-correcting codes.

Error correction with codes on graphs has been studied along two lines,
namely, by computing the average number of errors correctable with some
decoding algorithm by codes from a certain random ensemble of graph codes,
or by examining explicit code families whose construction involves graphs
with a large spectral gap.
The first direction originates in the works of Gallager
\cite{gal63} and Zyablov and Pinsker \cite{zya75} who showed
that random LDPC codes of
growing length can correct a nonvanishing fraction of errors.
Recently the decoding algorithm of \cite{zya75} was studied by
Burshtein \cite{bur08}
who derived an improved
estimate of the number of correctable errors compared to \cite{zya75}
and by Zyablov et al. \cite{zya09} who provided estimates of the number
of errors
under the assumption of local single error-correcting (Hamming) codes.
The second line of work, initiated in Tanner's paper \cite{tan81} and in
Sipser and Spielman's \cite{sip96},
pursues estimates of error correction with codes on regular graphs with a small second
eigenvalue and ensuing expansion properties.
Presently it is known that such codes under iterative decoding
can correct the number of errors equal to a half of the designed distance
of graph codes  \cite{bar05}. This estimate fits in a series of analogous results
for various ``concatenated'' coding schemes and has prompted a view
of graph codes as parallel concatenations of the local codes \cite{bar05}.
However, this result relies on certain restrictive assumptions
discussed below.

An extension of Tanner's construction
from graphs to hypergraphs was proposed
by Bilu and Hoory \cite{bil04} who showed that such codes (for high code rates)
can have minimum distance greater than the best known bipartite-graph constructions.
Interestingly, the codes considered in \cite{bil04} are a direct
extension of a construction in \cite{gal63} in the same way as
Tanner's graph codes extend LDPC codes.

As is well known, graphs with high expansion and random graphs share many
properties that can be used to prove estimates of error correction.
This similarity in the coding theory context
was emphasized in our recent work \cite{bar08b} where we showed
that ensembles of codes on random graphs
and explicit expander-like constructions share many common features
such as properties of the minimum distance and weight distribution.

Regarding the proportion of errors corrected by graph codes under
iterative decoding, we note one difference between
(generalized) LDPC codes on random graphs and explicit constructions based on
the graph spectrum. The explicit constructions based on regular graphs
depend on the difference between the largest and the second largest eigenvalue
of the graph (the ``spectral gap''). For this reason,
one is forced to rely on local codes with rather large minimum distance $d_0$,
for instance, $d_0$ greater than the square root of the degree $n$ of the graph.
Even though in the construction of \cite{sip96} and later works $n$ is kept
constant, this effectively rules out of consideration
local codes with small minimum distance such as the Hamming codes and the like.
The square root restriction is implied by the spectral gap of regular
bipartite graphs, and is the best possible owing to the Alon-Boppana
bound for graph spectra \cite{nil91}.
The purpose of the present work is to lift this limitation on the distance
$d_0$ by switching from
graphs with a large spectral gap to random graphs.

In this paper we obtain new estimates of the number of correctable errors
for random ensembles of bipartite-graph and hypergraph codes under iterative
decoding.
The first part of the paper is devoted to codes on regular
bipartite graphs. To construct long graph codes, we assume that the
degree of the graph is fixed and the number of vertices in both parts
approaches infinity.
Assuming that local constraint codes are used to correct 2 or more errors,
we show that almost all codes in the ensemble of graph codes
are capable of correcting all error patterns of weight that forms a constant
fraction of the code length.
This is a much less restrictive assumption on the local codes than
the one taken in earlier works on decoding of graph codes
\cite{zem01,bar05}. The proof of this result employs some ideas
of \cite{bar08b} introduced there for the analysis of the weight
distribution of graph codes.

We then observe that if the degree of the graph is allowed to increase
then graph codes with local codes of constant distance
do not correct a linearly growing number of errors under the proposed iterative decoding.
This motivates us to study graph codes with long local codes correcting a
growing number of errors that forms a fixed proportion of the degree. The results obtained
in this case parallel earlier theorems for product codes and graph codes
based on the spectral gap.

In the second part of the paper we establish similar results for codes
on hypergraphs, showing that a constant proportion of errors is corrected
by an iterative decoding algorithm that combines some ideas of
\cite{bar08b} with the results proved for codes on bipartite graphs in the
first part of the paper.
Constructing the code ensemble based on regular hypergraphs of a fixed
degree, we show that they contain codes capable of correcting a constant
proportion of errors. The proof involves no assumptions on the
distance of the local codes; in particular, we show that networks of
Hamming codes correct a fixed proportion of errors under iterative decoding.
This fact was previously proved by Tanner \cite{tan81} under the
assumption that the underlying graph is a tree. This assumption
is not needed in our results. As in the case of the graph ensemble, we also
perform the analysis of the decoding algorithm for the case of growing degree, finding
the proportion of errors correctable with hypergraph codes based on
long local codes.

This paper is dedicated to the memory of Ralf Koetter.
The first-named author discussed the problem of estimating the performance
of codes on graphs with Ralf in the beginning of 2004.
Ralf's idea at that point was to investigate the error correcting
capability of codes defined on some distance-regular graph, with local
constraints imposed at the vertices of the graph.
Presently it is understood that the setting most amenable to analysis
is that of codes defined on a regular bipartite graph.
Ralf himself made an initial attempt to analyze such codes in
a joint paper with Xiangyu Tang \cite{tan07}. The emphasis in \cite{tan07}
is on the estimation of the largest channel error rate tolerated by graph codes
under such decoding.
In the present paper, similarly
to \cite{sip96,zem01} and later works, the local codes are decoded up to their
correction radius guaranteed by the minimum distance.

\section{Code ensembles}
An $[N,K]$ binary linear code is a linear subspace of $\{0,1\}^N$ of dimension
$K.$ To construct an $[N,RN]$ binary linear graph code $C,$
consider an $n$-regular bipartite graph $G(V=V_1\cup V_2, E),$
where the set of vertices $V$ consists of two disjoint parts $V_1,V_2$
of size $m$ each, all the edges are of the form $(u,v), u\in V_1,v\in V_2,$
and the degree of every vertex $v$ in $V$ is $n.$ Let
$A[n,R_0n,d_0]$ be a linear binary code of length $n$ called the local code below.
We identify the coordinates of $C$ with the set $E$ and for a vertex $v \in V$
denote by
$\bfx(v)\in \{0,1\}^n$ the projection of a vector
$\bfx\in \{0,1\}^N, N=nm,$ on the edges incident to $v.$
A graph code $C(G)$ is defined as follows:
    \begin{equation}\label{eq:def}
      C=\{\bfx\in \{0,1\}^N:\; \forall_{v\in V}\bfx(v)\in A\}.
    \end{equation}
The ensemble of codes $\sG(A,m)$ is constructed by associating a code
$C(G)$ with a graph $G$ sampled from the set of graphs defined by a random
permutation on $N$ elements which establishes how the edges originating
in $V_1$ are connected to the vertices in $V_2.$

Generalizing this construction,
consider an $l$-partite $n$-regular uniform hypergraph $H=(V,E)$
i.e., a finite set $V=V_1\cup\dots \cup V_l,$ where $|V_1|=\dots=|V_l|=m,$
and a collection $E$
of $l$-subsets (hyperedges) of $V$ such that every $e\in E$
intersects each $V_i, 1\le i\le l$ by exactly one element and
each vertex $v\in V$ appears in exactly $n$ different subsets of $E$.
Aiming at constructing an $[N,RN]$ binary linear code $C$ by imposing local
constraints at the vertices, we again identify the coordinates
of $C$ with the (hyper)edges of $H$.
By definition, the code $C$ is formed of the vectors $\bfx$ that
satisfy condition (\ref{eq:def}) for every vertex in $V.$
The ensemble of codes $\sH(A,l,m)$ in this case is constructed by sampling a random hypergraph from the set of hypergraphs defined by $l-1$ independent
random permutations on $N$ elements. For $i=1,2,\dots, l-1,$ the $i$th permutation accounts for the placement
of edges between parts $V_1$ and $V_{i+1}$ of $H$. Of course, $\sH(A,l,m)$ becomes
$\sG(A,m)$ for $l=2$.

The following is known about the parameters of codes in the
graph and hypergraph ensembles considered here.
It is easy to see that the rate $R$ of the codes $C\in \sH(A,l,m)$ satisfies
$R \ge lR_0 -(l-1), l=2,3,\dots .$
Denote by $d(\sH)=d(\sH(A,l,m))$ the average value of the minimum distance
of codes in the hypergraph ensemble and let
  \begin{equation}\label{eq:distance}
   \delta=\delta(\sH)\triangleq \liminf_{N\to\infty}\frac{d(\sH)}N.
  \end{equation}
A way to bound the value of
$\delta$ below using the distribution of distances in the local code $A$
was suggested in \cite{bou99,len99}. More explicit results in this direction
were obtained in \cite{bar06,bar08b}. In particular, \cite{bar08b} shows
that $\delta(\sH)>0$ if the local distance $d_0$ satisfies $d_0>l/(l-1).$
For the bipartite graph ensemble $\sG(A,m)$ (i.e., for $l=2$) this implies that $d_0\ge 3,$
i.e., with high probability codes in ensemble are asymptotically good
(have nonvanishing rate and relative distance) when the local
codes correct one or more errors. For hypergraphs with $l=3$ or more
parts any local codes (without repeated vectors) account for an
asymptotically good ensemble.
An explicit lower bound for $\delta(\sH)$ that depends only on $l$ and
$d_0$ is given by \cite{bar08b}, see Theorem \ref{thm:distance} below.
For the case when $n$ is large and $d_0=\delta_0 n,$ a lower
estimate of $\delta(\sH)$ is given by the solution for $x$
of the following equation \cite[Cor.~6]{bar08b}:
  \begin{equation}\label{eq:delta}
  \frac{h(x)}{x} =\frac{l}{l-1}\frac{h(\delta_0)}{\delta_0}.
  \end{equation}

Finally, if the local codes are chosen randomly as opposed to a fixed
code $A$ used at every vertex of $H,$ then the codes in the (hyper)graph
ensemble match the best known linear codes, i.e., reach the asymptotic
Gilbert-Varshamov bound on the minimum distance \cite{bar08b}.

\begin{figure*}[tH]
\centering
\includegraphics[scale=0.8]{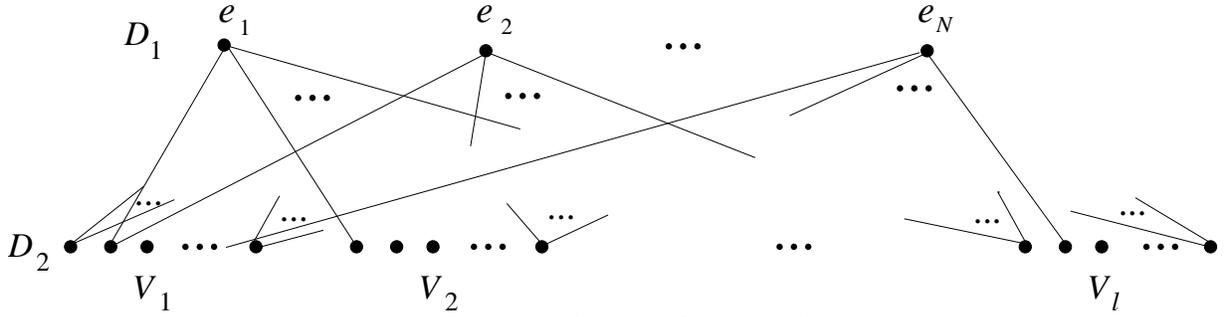}
\vspace*{-9pt}
\caption{Alternate construction of the hypergraph code: The set
$D_1=\{e_1,\dots,e_N\},$ where $\deg(e_i)=l$ for all $i$,
represents the coordinates of the code (hyperedges of $H$);
the sets $V_1,\dots, V_l,$ where $|V_j|=m$ for all $j$, represent the vertices of
the hypergraph $H$.
Each vertex $v_{i,j}, 1\le i\le l, 1\le j\le m$ carries a codeword of the
local code $A$ of length $n$. }\label{fig1}
\end{figure*}

{\em Remarks.} 1. An equivalent description of the bipartite code
ensemble is obtained
by considering an edge-vertex incidence graph of the graph $G(V,E)$, i.e., a
bipartite graph $D=(D_1\cup D_2, \bar E)$ where
$D_1=E, D_2=V_1\cup V_2,$ each vertex in $D_1$ is connected to one vertex
in $V_1$ and to one vertex in $V_2,$ and there are no other edges
in $\bar E.$ Thus, for all $v\in D_1, \deg(v)=2$ and for all $v\in D_2,$
$\deg(v)=n.$ The local code constrains are imposed on the vertices in
$D_2$.
By increasing the number of parts in $D_2$ from two to $l$, we then
obtain the hypergraph codes defined above. This gives an alternate description of
the hypergraph code presented in Fig.~\ref{fig1}.

The ensemble of hypergraph codes
with local constraints given by single parity-check codes was introduced
by Gallager \cite[p.12]{gal63}. The proportion of errors correctable with these codes using the so-called ``flipping'' algorithm
was estimated in \cite{zya75}. Several generalizations
of this ensemble were studied in \cite{bil04,bar08b}.

2. The derivations of this paper are not specific to binary codes:
any local linear codes such as Reed-Solomon codes can be used in the
construction with no conceptual changes to the analysis and the conclusions.

\section{Decoding algorithms for graph (Generalized LDPC) codes}
Even though the ensemble $\sG(A,m)$ forms a particular case of the ensemble
$\sH(A,l,m),$ in our analysis we employ different decoding algorithms for the
cases $l=2$ and $l\ge 3.$ The reason for this is that edge-oriented
procedures commonly used for bipartite-graph codes do not generalize
well to hypergraphs.

\emph{A. Decoding for the ensemble $\sG(A,m)$.}
In our estimates of the number of correctable
errors for the ensemble $\sG$ we rely
upon the algorithm of \cite{zem01} which iterates between decoding
all the vertices in parts $V_1$ and $V_2$ in parallel using some
decoding algorithm of the code $A$. Let $C\in \sG(A,m)$ be a code.
For the ease of analysis we
assume that the local codes are decoded to correct up to $t$ errors,
where $t\ge 0$ is an integer that satisfies $2t+1\le d_0$ and $d_0$ is
the distance of the code $A.$
Formally, define a mapping $\psi_{A,t}:\{0,1\}^n\to \{0,1\}^n$ such that
$\psi_{A,t}(\bfz)=\bfx\in A$ if $\bfx$ is the unique codeword that
satisfies $d(\bfz,\bfx)\le t$ and $\psi_{A,t}(\bfz)=\bfz$ otherwise.
Let $\bfy^{(i)}$ be the estimate of the transmitted vector before the $i$th
iteration, $i\ge 1,$ where $\bfy=\bfy^{(1)}$ is the received vector.
The next steps are repeated for a certain number
of iterations.

{\em Algorithm I} $(\bfy^{(1)})$\\[1mm]
 \hspace*{3mm} $\bullet$ $i$ odd: for all $v\in V_1$ put $\bfy^{(i+1)}(v)=\psi_{A,t}(\bfy^{(i)}(v));$\\
  \hspace*{3mm} $\bullet$ $i$ even: for all $v\in V_2$ put $\bfy^{(i+1)}(v)=\psi_{A,t}(\bfy^{(i)}(v)).$

\remove{Then in odd iterations the algorithm performs decoding of all the local
codes in part $V_1,$ subsequently replacing the current estimate
of the transmitted vector by the result of this decoding. If a
particular vertex fails to find a codeword within distance $t$
from its current estimate, then no changes are made from the current estimate.
In even iterations this procedure is repeated for the part $V_2.$
Below we call this decoding Algorithm I.}

\vspace*{2mm}\emph{B. Decoding for the ensemble $\sH(A,l,m)$.}
For the hypergraph ensemble $\sH(A,l,m)$ we use the decoding
algorithm proposed in \cite{bar08b}. It proves to be the best choice in terms
of the number of correctable errors among several possible
algorithms for these codes such as the one in \cite{bil04}
and procedures analogous to Algorithm I above.

Let $C\in \sH(A,l,m)$ be a code and let $H(V,E), V=V_1\cup\dots\cup V_l$
be the graph associated with it.
For every $i=1,2\dots,l$ we will define an $i$-th subprocedure that
decodes the local code $A$ on every vertex in the part $V_i.$ Suppose
that a vector
$\bfu\in\{0,1\}^N$ is associated with the edges $e\in E.$
Let $v_{i,1},\dots,v_{i,m}$ be the vertices in the part $V_i$ of $H$
and let $\bfu_{i,1}=\bfu(v_{i,1}),\dots,\bfu_{i,m}=\bfu(v_{i,m})$ be the $m$ subvectors
obtained from $\bfu$ upon permuting its coordinates according to the
order of edges in $V_i$ and projecting it on the vertices
$v_{i,1},\dots,v_{i,m}.$ In other words, the vector
$(\bfu_{i,1},\dots,\bfu_{i,m})$ is obtained from $\bfu$ using the
permutation that establishes edge connections between parts $V_1$ and $V_i.$
The $i$th subprocedure replaces the vector
$(\bfu_{i,1},\dots,\bfu_{i,m})$ with the vector
$(\psi_{A,t}(\bfu_{i,1}),\dots,\psi_{A,t}(\bfu_{i,m})).$

The algorithm proceeds in iterations.
Let $\bfy\in\{0,1\}^N$ be the received vector.
Denote by
$Y^{(j)}$ the set of estimates of the transmitted
codeword (i.e., the set of $N$-vectors) stored at the vertices of $H$
before the $j$th iteration $j=1,2,\dots.$ After each iteration, this set
is formed as the union of the vectors obtained upon decoding of the
vertices in the $i$th part, $i=1,\dots, l.$
Decoding begins with setting $Y^{(1)}=\{\bfy\}.$
After the first iteration we obtain $l$ potentially different vectors
(one for each subprocedure)
which form the current estimates of the transmitted vector.
These vectors form the sets $Y_i^{(2)}, i=1,\dots, l.$
In the next iteration each subprocedure will have to be applied to each
of the $l$ outcomes of the preceding iteration.
Proceeding in this way, we observe that $|Y_i^{(j)}|\le l^{j-1}.$

This algorithm, called Algorithm II below, will only be applied for
a constant number $s$ of iterations until we can guarantee that at
least one subprocedure has reduced the number of errors to a specified
proportion, say from $\gamma_0 N$ to some $\gamma_1 N, \gamma_1<\gamma_0.$
We then let another algorithm take over and decode all the $l^s$ candidates.
Any low-complexity
decoder of graph codes that removes an arbitrarily small positive fraction
of errors $\gamma_1$ will do at this stage. This is because taking the proportion of
errors from $\gamma_0$ to $\gamma_1>0$ can be accomplished in a
constant number $s$ of steps, so the number of candidates that this decoder has
to handle is at most $l^{s}$ and does not depend on $N.$

For the case of local codes correcting $t\ge 2$ errors
we let this algorithm to be the decoding algorithm of bipartite-graph
codes (Algorithm I), making sure that $\gamma_1$ is below the proportion
of errors that are necessarily corrected by this algorithm for the
ensemble $\sG(A,m).$ This is possible because, leaving any two parts of
the original hypergraph $H$ to form a bipartite graph $G$, we obtain a
random code from the ensemble $\sG(A,m)$ which with high probability
(over the ensemble) will
remove all the residual errors from at least one candidate estimate.
For $t=1$ this approach fails for the reasons discussed in the next section,
so we resort to a procedure in \cite{zya09} that corrects a small linear fraction
of errors for single-error-correcting Hamming codes.

Upon performing the described procedure we obtain a list of at most $l^s$
candidate codewords of the code $C.$ The final decoding result is 
found by choosing the codeword from this list closest to $\bfy$ by 
the Hamming distance. 

Though the last step of the decoding algorithm described is different
from \cite{bar08b}, the main idea is similar to that paper,
so we refer to it for a more detailed description and a
discussion of the algorithm.

\section{Number of correctable errors for the ensemble $\sG(A,m)$}

Let $C\in\sG(A,m)$ be a code and let $G(V,E)$ be the graph
associated with it. For a given subset of vertices
$S\subset V_i,i=1,2$ and a vertex $v$ denote by $\deg_S(v)$ the
number of edges between $v$ and $S.$
Let $T_r(S)=\{v\in V: \deg_S(v)\ge r+1\},$
where $r\in\{0,\dots,n-1\}$ is an integer.

Below $h(\bfz)$ denotes  the entropy of the probability vector
$\bfz\in \reals^{n+1}$. 
In the particular case of $n=1$ we write $h(z)$ instead of $h(z,1-z).$

Let $t\ge 0$ be any integer such that $2t+1\le d_0.$
The calculation in this section is based on the following simple observation.
\begin{proposition}\label{prop:decoding}
Suppose that for all
$S\subset V_i, i=1,2, |S|\le \sigma m, \sigma\in(0,1),$ there exists $\epsilon>0$
such that $|T_t(S)|\le|S|-\epsilon m.$ Then any $\sigma t m=\sigma t(N/n)$
errors will be corrected by Algorithm I in $O(\log m)$ iterations.
\end{proposition}
\begin{IEEEproof}
Suppose that no more than $\sigma t m$ errors occurred in the channel.
Let $S_i$ be the set of vertices that are decoded incorrectly in iteration $i$
of Algorithm I.
The assumption of the proposition implies that $|S_{i+1}|\le |S_i|(1-\epsilon/\sigma),$
so $O(\log m)$ iterations suffice to remove all the errors.
\end{IEEEproof}
Define
\begin{align}
     F_{n,t}(\sigma)&=h(\sigma)-\sigma n\log x\nonumber\\
         &+\sigma \log\sum_{i=t+1}^n\binom ni x^i
      +(1-\sigma)\log\sum_{i=0}^t\binom ni x^i, \label{eq:m}
   \end{align}
where $x>0$ is found from the equation
   \begin{equation}\label{eq:x}
     \sum_{i=0}^t\sum_{j=t+1}^n\binom ni\binom nj(\sigma(n-j)-i(1-\sigma))
   x^{i+j-t-1}=0.
   \end{equation}
Let $\cZ_n=\{\bfz\in[0,1]^{n+1}:\, \sum_{i=0}^nz_i=1\}$ be
the $(n+1)$-dimensional probability simplex.

The main result of this section is given by the next theorem.
\begin{theorem}\label{thm:bg} Let $A[n,R_0n, d_0]$ be the local code,
let $m\to\infty$, and let $2\le t<d_0/2.$
All codes in the ensemble $\sG(A,m)$
except for an
exponentially small (in $N$) proportion of them correct any combination of errors
of weight  $\sigma tm$ in $O(\log m)$ iterations
of Algorithm I, where $0<\sigma<\sigma_0$ and $\sigma_0$ is the
smallest positive root of the equation
    $$
       F_{n,t}(\sigma)=(n-1)h(\sigma).
    $$
\end{theorem}
{\em Remark.} The case of local codes with $t=1$ is excluded from this
theorem because $G$ with high probability contains a large number of
4-cycles, which means that correcting single error at every vertex
does not ensure overall convergence of the decoding. Indeed, if
two vertices are affected by two errors each, and the corresponding 4 edges
form a cycle, then the decoder will loop indefinitely without approaching
the correct decision. The theorem
is still valid in this case, but gives $\sigma_0=0.$

\begin{IEEEproof} We need to verify the assumption of Proposition
\ref{prop:decoding}.
Let $S\subset V_1,|S|=\sigma m$ and let $m_i=|\{v\in V_2: \deg_{S}(v)=i\}|,
i=1,\dots, n.$  Clearly,
   $$
        \sum_{i=1}^n m_i\le m,\; \sum_{i=t+1}^n m_i=|T_t(S)|,\;
\sum_{i=1}^n im_i=|S|n.
   $$
Let us compute the probability (over the choice of $G$) that
$|T_t(S)|\ge(\sigma-\epsilon)m.$ Let ${\boldsymbol\mu}=(m_1,\dots,m_n)$
be a vector with nonnegative integer components, let
   \begin{multline*}
    M_\epsilon(t,\sigma)=\{{\boldsymbol\mu}: \;\sum_{i=1}^n m_i\le m,\,
                    \sum_{i=1}^n im_i=\sigma N,\;\\[-1mm]
             \sum_{i=t+1}^n m_i\ge
     (\sigma-\epsilon) m\},
   \end{multline*}
and let $\binom m{\boldsymbol\mu}$ denote the number of choices of subsets
of size $m_1,\dots, m_n$ out of a set of size $m$. We have
  \begin{equation}\label{eq:pT}
   P(|T_t(S)|\ge|S|-\epsilon m)=\frac1{\binom N{\sigma N}}
        \sum_{{\boldsymbol\mu}\in M_\epsilon(t,\sigma)}\binom m{\boldsymbol\mu}\prod_{i=1}^n\binom ni^{m_i}.
  \end{equation}
Let $\sL_1(s)$ denote the event that $V_1$ contains a subset
$S, |S|=s$ for which $|T_t(S)|\ge |S|-\epsilon m.$ We have
  $$
  P(\sL_1(\sigma m))\le \binom m{\sigma m}P(|T_t(S)|\ge|S|-\epsilon m)
  $$
and
  $$
  P\Big( \bigcup\limits_{i=1}^{\sigma m}\sL_1(i)\Big)\le
    m P(\sL_1(\sigma m)).
  $$
Denote by $\sL_2(\sigma)$ an analogous event with respect to $V_2.$
Then
  \begin{equation}\label{eq:P}
  P\Big(\bigcup\limits_{i=1}^{\sigma m}(\sL_1(i)\cup\sL_2(i))\Big)
    \le  \frac{2m\binom m{\sigma m}}{\binom N{\sigma N}}
     \!\sum_{{\boldsymbol\mu}\in M_\epsilon(t,\sigma)}\!\binom m{\boldsymbol\mu}\prod_{i=1}^n\binom ni^{m_i}.
  \end{equation}
Letting $L$ to be the logarithm of the left-hand side divided by $m$ and
omitting $o_m(1)$ terms,
we obtain the estimate $L\le  n^{-1} \bar F_{n,t}(\sigma),$ where
   $$
  \bar F_{n,t}(\sigma)= -(n-1) h(\sigma)
+\max_{\bfz\in \sM'_\epsilon(t,\sigma)}
   \Big(h(\bfz)+ \sum_{i=1}^n z_i\log\binom ni\Big),
   $$
where
   $$
 \sM'_\epsilon(t,\sigma)=\Big\{\bfz\in\cZ_n:
\sum_{i=1}^n iz_i=\sigma n, \,\sum_{i=t+1}^n z_i\ge\sigma-\epsilon\Big\}
   $$
and $z_i=m_i/m, z_0=(m-\sum m_i)/m.$

The rest of the proof is concerned with the evaluation of the above maximum.
Define
  \begin{equation}\label{eq:g}
     g(\bfz)=h(\bfz)+ \sum_{i=1}^n z_i\log\binom ni
  \end{equation}
  $$
 \bar\sigma=\sup\{\sigma>0: \bar F_{n,t}(y)<0 \text{ for all } 0\le y<\sigma\}.
  $$
As long as $\sigma<\bar\sigma,$ the probability of not being able to correct
$\sigma tm$ errors with a random code from the considered ensemble approaches zero.
Thus, we need to
find the maximum $\max_{\bfz\in \sM'_\epsilon(t,\sigma)}g(\bfz)$
for all $\sigma\in[0,\bar\sigma).$
The proof will be accomplished in the next three steps.
Since $\epsilon$ will be assumed arbitrarily small, we will omit it from our
considerations and write $\sM'$ instead of $\sM'_\epsilon$.

1. We find the point $\bfz^\ast$ that gives the maximum of $g(\bfz)$ without the
constraint $\sum_{i=t+1}^n z_i\ge \sigma.$

2. Next we show that for $0\le\sigma<\bar\sigma,$ the point $\bfz^\ast\not \in
\sM',$ and therefore the maximum over $\sM'$ is attained on the boundary, i.e.,
we can replace $\sM'$ with
   $$
   \sM(t,\sigma)=\Big\{\bfz\in\cZ_n:
\sum_{i=1}^n iz_i=\sigma n, \,\sum_{i=t+1}^n z_i=\sigma\Big\} .
   $$

3. Finally we compute the value of the maximum.

\vspace*{.05in}Step 1. Without the constraint $\sum_{i=t+1}^n z_i\ge \sigma$
the maximum is easily computed.
Indeed, the proportion of edges incident to the vertices in $S$
out of the $N$ edges of $G$ is $\sigma,$
so the fraction of vertices with $S$-degree $i$ should be close to
$z_i^\ast(\sigma)=\binom ni \sigma^i (1-\sigma)^{n-i}.$ Thus, the coordinates of
the maximizing point
$\bfz^\ast=\bfz^\ast(\sigma)$ are $z_i^\ast, i=1,\dots,n; z_0=1-\sum_iz_i^\ast,$ and
 \begin{equation*}
  g(\bfz^\ast)=n h(\sigma).
 \end{equation*}
Slightly more formally, note that $\bfz^\ast$ is the unique stationary
point of the function $g(\bfz)$, and that this function is strictly
concave in $\bfz.$  Therefore, $\bfz^\ast$ is a unique maximum of
$g(\bfz)$ on $\cZ_n$, and the function  $g(\bfz)$ grows in the
direction $\bfz^\ast-\bfz$ for any $\bfz\in \cZ_n.$

\vspace*{.05in}Step 2. Suppose that $0\le\sigma\le\bar\sigma.$
Observe that $p(\sigma)\triangleq\sum_{i=t+1}^n z_i^\ast=P(X\ge t+1),$
where $X$ is a $(\sigma,1-\sigma)$ binomial random variable.
This probability is monotone increasing on $\sigma$ for $\sigma\in[0,1],$
and $p(0)=p'(0)=0.$
Thus for $\sigma\in [0,\alpha)$ where $\alpha$ is the smallest positive root of
$\sum_{i=t+1}^n z_i^\ast(\sigma)=\sigma,$  we have
   $$
    \sum_{i=t+1}^n z_i^\ast=\sum_{i=t+1}^n \binom ni\sigma^i(1-\sigma)^{n-i}
       <\sigma,
   $$
and so the point $\bfz^\ast(\sigma)\not\in \sM'(t,\sigma).$
Our claim will follow if we show that $\bar \sigma<\alpha.$ This is indeed
the case because for $0\le\sigma<\bar \sigma,$
  $$
     \max_{\bfz\in \sM'(t,\sigma)} g(\bfz^\ast(\sigma))<(n-1)h(\sigma).
  $$
On the other hand, $g(\bfz^\ast(\alpha))=nh(\alpha).$
This establishes that the maximum of $g(\bfz)$ on $\bfz\in \sM'$ is attained
on the hyperplane $\sum_{i=t+1}^n z_i=\sigma.$

\vspace*{.05in}Step 3.
To compute the maximum of $g(\bfz)$ on $\bfz,$ let us form the
Lagrangian
 \  \begin{multline*}
    U(\bfz,\tau_1,\tau_2)=h(\bfz)+\sum_{i=1}^n z_i\log\binom ni\\+
    \tau_1\Big(\sum_{i=1}^n iz_i-\sigma
    n\Big)+\tau_2\Big(\sum_{i=t+1}^n z_i-\sigma\Big).
  \end{multline*}
Setting $\nabla U=0$ and $\tau_1=\log x, \tau_2=\log y,$ we find that
  $$ z_i=\begin{cases} {\displaystyle \binom ni
  x^iD}&\quad\text{if }0\le i\le t\\[.2in] {\displaystyle \binom
  ni yx^iD}&\quad\text{if }t<i\le n,
     \end{cases}
     $$ where we have denoted
     $$ D=\Big[\sum_{i=0}^t\binom ni x^i+y\sum_{i=t+1}^n\binom ni x^i\Big]^{-1}.
     $$
Adding these equations together, we find conditions for $x$
and $y$:
   $$ \sigma=Dy\sum_{i=t+1}^n\binom ni x^i
   $$
   $$ \sigma n=D\Big(\sum_{i=0}^ti\binom ni x^i+y\sum_{i=t+1}^ni\binom
   ni x^i\Big).
   $$
Once $y$ is eliminated from the last two equations, we obtain
the condition (\ref{eq:x}) for $x.$ Finally, substituting the found
values of $z_i,i=1,\dots,n$ into $g(\bfz),$ we find that
the maximum evaluates to the expression
$F_{n,t}(\sigma)$ given in (\ref{eq:m}) (and therefore, $\bar\sigma=\sigma_0).$
Since we seek to obtain a
value $L<0,$ the boundary condition for the proportion of correctable
errors is obtained by setting $L=0.$ This concludes the proof.  \end{IEEEproof}

\vspace*{.1in}
\begin{example} Using Theorem \ref{thm:bg} together with (\ref{eq:m})
we can compute the proportion of errors corrected by codes in the
ensemble $\sG(A,m), m\to \infty$ for several choices of the local code $A$.
For instance, taking $A$ to be the binary Golay code of length $n=23$
we find $\sigma_0\approx0.0048586$ and therefore, the proportion of
correctable errors is $\frac{\sigma_0 t}n\approx0.00063.$ Similarly,
for the 2-error-correcting $[n=31,k=21]$ BCH code we find
$\sigma_0\approx 0.000035$ and $\frac{\sigma_0 t}n\approx 0.0000023.$
\end{example}

\vspace*{.1in}
To underscore similarities with the results obtained for
product codes and their later variations including graph codes
(e.g., \cite{zem01}) we compute the proportion of errors correctable with
codes from the ensemble $\sG(A,m)$ in the case of large $n$.
\begin{proposition} Let $t=\tau n.$ Then the ensemble $\sG(A,m)$ contains
codes that correct  $\sigma\tau N$ errors for any $\sigma\le\sigma_0,$ where
$\sigma_0$ is given by
   \begin{gather*}
    \sigma_0=\sup\Big\{\sigma>0: \forall_{0<x<\sigma}\;(1-x)
         h\Big(\frac{x(1-\tau)}
     {1-x}\Big)\\+x h(\tau)+\varepsilon_n<h(x)\Big\}
   \end{gather*}
where $\varepsilon_n=(1+\log n)/n.$
\end{proposition}
\begin{IEEEproof} Referring to the notation of the previous proof,
let us evaluate the asymptotic behavior
of the exponent $L$ of the probability in (\ref{eq:P}).
  Since $h(\bfz)\le\log n,$ we have
  \begin{multline*}
    n^{-1}\bar F_{n,t}(\sigma)\le -h(\sigma)+n^{-1}
\max_{\bfz\in \sM(\tau n,\sigma)} \sum_{i=0}^nz_i\log\binom ni
   \\+n^{-1}(1+\log n).
  \end{multline*}
Next,
  $$
    \frac 1n\sum_{i=0}^nz_i\log\binom ni\le \sum_i z_ih\Big(\frac in\Big)
  $$
\begin{align*}
   &=(1-\sigma)\sum_{i=0}^t\frac{z_i}{1-\sigma}h\Big(\frac in\Big)
  +\sigma \sum_{i=t+1}^n \frac {z_i}{\sigma}h\Big(\frac in\Big)\\
   &\le (1-\sigma)h\Big(\frac{\sum_{i=1}^t iz_i}{(1-\sigma)n}\Big)
     +\sigma h\Big(\frac{\sum_{i=t+1}^n iz_i}{\sigma n}\Big).
 \end{align*}
Let $y=n^{-1}\sum_{i=t+1}^n iz_i,$ then for any $\bfz\in\sM(\tau n,\sigma)$ we have
  $$
\frac 1n\sum_{i=0}^nz_i\log\binom ni\le \max_{\tau\sigma\le y\le\sigma}\Big\{
      (1-\sigma)h\Big(\frac{\sigma-y}{1-\sigma}\Big)
     +\sigma h\Big(\frac{y}{\sigma}\Big)   \Big\}.
  $$
The function on the right-hand side of this inequality is concave. Its
global maximum equals $h(\sigma)$ and is attained for $y=\sigma^2.$
Thus, assuming that $\sigma<\tau,$
we conclude that the constrained maximum occurs for $y=\tau \sigma,$ which gives
the following bound on $n^{-1}\bar F_{n,t}(\sigma):$
   $$
   n^{-1}\bar F_{n,t}(\sigma)\le -h(\sigma)+(1-\sigma)h\Big(\frac{\sigma(1-\tau)}{1-\sigma}\Big)
  +\sigma h(\tau)+\varepsilon_n.
   $$
As long as the right-hand side of the this inequality is negative,
the previous proof implies that the code corrects all errors of multiplicity
up to $\sigma\tau N.$
\end{IEEEproof}

From the expression of this proposition we observe that (as $n\to\infty$)
the value of $\sigma_0$ approaches $\tau,$
so the ensemble $\sG$ contains codes that correct up to a $\tau^2$
proportion of errors, where $\tau n=d_0/2$ is the error-correcting capability
of the code $A.$ This result parallels the product bound on the error-correcting
radius of direct product codes.
As in the case of product and expander codes (e.g., \cite{bar05}), the
proportion of correctable errors can be improved from $\tau^2=(d_0/(2n))^2$
by using a more powerful decoding algorithm.

\section{Number of correctable errors for the ensemble $\sH(A,l,m)$}
\label{sect:hypergraph-decoding}
In this section we first state a sufficient condition for the existence
of at least one subprocedure within each step of Algorithm II that reduces the
number of errors, and then perform the analysis of random hypergraphs
to show that with high probability this condition is satisfied. Overall
this will show that the number of errors in at least one of the candidates
in the list generated after a few iterations is reduced to a desired
level. 

Denote by $E(v)$ the set of edges incident to a vertex $v\in V.$
Let $C\in\sH(A,l,m)$ be a code and let $H(V,E)$ be its associated graph.
Let $\cE\subset E$ be the set of errors at the start of some iteration
of the algorithm. The next set of arguments will refer to this iteration.
Let $G_i=\{v\in V_i: |E(v)\cap \cE|\le t\}$ be the set of vertices such
that each of them
is incident to no more than $t$ edges from $\cE$ (such errors will be
corrected upon one decoding). Let
$B_i=\{v\in V_i: |E(v)\cap \cE|\ge d_0-t\}$ be the set of vertices that
can introduce errors after one decoding iteration. Note that each of such
vertices introduces at most $t$ errors.

The main condition for successful decoding is given in the next lemma.
\begin{lemma}\label{lemma:red} Assume that for every
$\cE\subset E, |\cE|\le \gamma N$ there exists
$i=i(\cE), 1\le i\le l$ such that $|\cE(G_i)|\ge t |B_i|+\epsilon N,$ where
$\cE(G_i)$ is the set of edges of $\cE$ incident to the vertices of $G_i$ and
$\epsilon>0.$  Then for any $0<\beta<\gamma,$
Algorithm II will reduce any $\gamma N$ errors in the received vector to
at most $\beta N$ errors in $c(\beta,\gamma,\epsilon),$ iterations where
$c$ is a constant independent of $N$.
\end{lemma}
\begin{IEEEproof} We need to prove that at least one of the subprocedures
will find a vector with no more than $\beta N$ errors after a
constant number of iterations. In any given iteration by the assumption of
 the lemma
there exists a component $V_i$ for which the $i$th subprocedure will
decrease the count of errors by $|\cE(G_i)|-tB_i\ge
\epsilon N.$ Thus, in each iteration
there exists a subprocedure that reduces the number of errors by
a positive fraction. \end{IEEEproof}

Next we show that the assumption of Lemma \ref{lemma:red} holds
with high probability over the ensemble. Consider the function
        \begin{multline*}\label{eq:barF}
  \tilde F_{n,t}(\gamma)= \max_{\bfz\in \sM(t,\gamma)}
   \Big(h(\bfz)+ \sum_{i=0}^n z_i\log\binom ni\Big),
   \end{multline*}
where in this section the region $\sM(t,\gamma)$ will be as follows:
      \begin{equation}\label{eq:mts}
    \sM(t,\gamma)=\Big\{\bfz\in \cZ_n: \sum_{i=1}^n iz_i=\gamma n,
         \sum_{i=1}^t iz_i=\sum_{i=d_0-t}^n tz_i\Big\}.
     \end{equation}
\begin{lemma}\label{prop:hg} Let $m\to\infty$ and let
    \begin{equation}\label{eq:gast}
    \gamma_0=\sup\{x>0: \forall_{0<\gamma\le x}\,
   (l/n) \tilde F_{n,t}(\gamma)<(l-1)h(\gamma)\}.
    \end{equation}
A hypergraph from the ensemble of $l$-partite uniform
$n$-regular hypergraphs with probability $1-2^{-\Omega(N)}$ has the
property that for all $\cE\subset E, |\cE|<\gamma_0 N,$ and some $\epsilon>0$,
the inequality $|\cE(G_i)|\ge t |B_i|+\epsilon N$ holds for at least one
$i\in\{1,\dots, l\}.$
\end{lemma}
\begin{IEEEproof} Let $\cE\subset E, |\cE|=\gamma N.$
Let $m_i=|\{v\in V_1: |E(v)\cap \cE|=i\}|, i=1,\dots,n.$ Clearly
$|\cE(G_1)|=\sum_{i=0}^t i m_i$ and $|B_1|=\sum_{i=d_0-t}^n m_i.$
We have
   \begin{multline*}
    p\triangleq P(|\cE(G_i)|\le t |B_i|+\epsilon N)\\=\frac 1{\binom N{\gamma N}}
      \sum_{{\boldsymbol\mu}\in M_\epsilon(t,\gamma)}
         \binom m{{\boldsymbol\mu}}\prod_{i=0}^n\binom ni^{m_i},
   \end{multline*}
where $\boldsymbol\mu=\{m_1,\dots,m_n\},$
  \begin{multline*}
   M_\epsilon(t,\gamma)=\{{\boldsymbol\mu}\in (\integers_+\cup0)^{n}:
      \sum_{i=1}^n m_i\le m, \\[-2mm]\sum_{i=1}^n im_i=\gamma N,
       \sum_{i=1}^t im_i\le\sum_{i=d_0-t}^n t m_i+\epsilon N\}.
  \end{multline*}
Denote by $\sL(\cE)$ the event that for a given subset $\cE\subset E,
|\cE|=\gamma N$ no part $V_i$ of $H$ satisfies the assumption of
Lemma \ref{lemma:red}. Then $P(\sL(\cE))=p^l$ and
  $$
    P\{\exists \cE:\; (|\cE|\le \gamma N)\wedge (\sL(\cE))\}
   \le N \binom N{\gamma N} p^l.
  $$
Letting $L$ to be the logarithm of the left-hand side of this inequality
divided by $N$ and omitting $o_N(1)$ terms, we obtain
  \begin{equation}\label{eq:L}
    L\le -(l-1)h(\gamma)+\frac ln\max_{\bfz\in \sM'(t,\gamma)} g(\bfz),
  \end{equation}
where $g(\bfz)$ is defined in (\ref{eq:g}),
  $$
   \sM'(t,\gamma)=\{\bfz\in \cZ_n: \sum_{i=1}^n iz_i=\gamma n,\,
     \sum_{i=1}^t iz_i\le \!\!\sum_{i=d_0-t}^n tz_i\}
  $$
and $z_i=m_i/m$ (as in the previous section, we have omitted $\epsilon$
which can be made arbitrarily small).

The proof will be complete if we show that the optimization region
$\sM'$ can be replaced by $\sM.$ For that we follow the logic of the second part
of the proof of Theorem \ref{thm:bg}. As before, the maximum of $g(\bfz)$ without
the constraint $\sum_{i=1}^t iz_i\le \sum_{i=d_0-t}^n tz_i$ is attained
at the point $\bfz^\ast(\gamma)=(z_0^\ast,z_1^\ast,\dots,z_n^\ast)\in \cZ_n,$  where
   $$
     z_i^\ast=z_i^\ast(\gamma)= \binom ni\gamma^i(1-\gamma)^{n-i}, \quad
   i=1,\dots, n.
   $$
We need to show that as long as $0\le\gamma <\gamma_0,$ the point
$\bfz^\ast\not\in\sM'(t,\gamma).$ By concavity of the objective function
and the optimization region, this will imply that the maximum is on the boundary.
As before, it is possible to show that in the neighborhood of $\gamma=0,$
  $$
   \sum_{i=1}^t iz_i^\ast> \sum_{i=d_0-t}^n tz_i^\ast.
  $$
and thus for $\gamma<\beta,$ where $\beta$ is the smallest positive root
 of $\sum_{i=1}^t iz_i^\ast=\sum_{i=d_0-t}^n tz_i^\ast,$ the point
$\bfz^\ast(\gamma)\not\in \sM'(t,\gamma).$ Let
  $$
  \bar\gamma=\sup\{\gamma:\;\forall\, {0<x<\gamma}, \;\text{rhs of (\ref{eq:L})}
     <0\}.
  $$
We note that for all $\gamma\le\bar\gamma,$
  $$
    \max_{\bfz\in \sM'(t,\sigma)} g(\bfz)<(l-1)nh(\gamma).
  $$
On the other hand, $g(\bfz^\ast(\beta))=n h(\beta).$ This implies that
$\bar\gamma<\beta,$ and so for all $\gamma<\bar\gamma,$ the point
$\bfz^\ast(\gamma)\not\in \sM'(t,\gamma).$ Thus the region
$\sM'$ in the maximization can be replaced with $\sM$ (and $\bar\gamma=\gamma_0$).
\end{IEEEproof}

This lemma establishes that
the number of errors in at least one of the candidates
in the list generated after a few iterations is reduced to a desired
level. After that the residual errors can be removed by another procedure
as described above. In this situation we say that the errors are correctable
by Algorithm II, without explicitly mentioning the second stage.

%
In the next theorem, which is the main result of this section, $\delta$
refers to the lower estimate of the average relative distance of the 
hypergraph code ensemble $\sH$ from Theorem  \ref{thm:distance} below.
\begin{theorem} Let $t\ge 2$ be the number of errors correctable by the local code
$A.$ 
Algorithm II corrects any combination of up to $N(\min(\gamma_0,\delta/2))$
errors for any code $C\in \sH(A,l,m)$ except for a proportion of codes
that declines exponentially with the code length $N=nm, m\to\infty.$
\end{theorem}
\begin{IEEEproof} With high probability over the ensemble of hypergraphs
considered, for a given hypergraph $H(V,E)$
a constant number $s$  of iterations of the algorithm will
decrease the weight of error from $\gamma_0 N$ to any given positive
proportion $\beta$ for at least one of the $l^s$ candidates in the list
$Y^{(s+1)}_1.$ Take $\beta=\sigma_0,$ where $\sigma_0$ is the quantity
given by Theorem \ref{thm:bg}. Next consider the bipartite
graph $G(V_G=V_1\cup V_2, E_G)$ where $V_1,V_2$ are the parts of $H$
and where $(v_1,v_2)\in E_G$ if $v_1,v_2\in e$ for some edge $e\in E.$
By the previous section, with high probability these $\sigma_0 N$
errors can be corrected with $O(\log m)$ iterations of Algorithm I.
Finally, the correct codevector will be selected from the list of candidates
because the proportion of errors is assumed not to exceed $N\delta/2.$
\end{IEEEproof}

The complexity of this decoding is $O(N\log N)$ where the implicit constant
depends on the code $A.$

In the following theorem we extend the results of this section to the case
of $A$ being a perfect single-error correcting Hamming code of length
$n=2^r-1$ for some $r=3,4,\dots.$
In this case the maximum on $\bfz$ in the above proof can be computed in a
closed form. As remarked above, in this case
in the last part of the error correction procedure
we use the decoding algorithm of \cite{zya09} to remove residual errors
from the candidate vectors.
\begin{theorem} Suppose that the local codes $A$ are taken to be
one-error-correcting Hamming codes and let
$\delta=\delta(\sH)$ be the relative average distance (\ref{eq:distance})
of the ensemble $\sH(A,l,m).$
Then almost all codes in the ensemble $\sH(A,l,m)$ can be decoded
to correct $N\min(\gamma_0,\delta/2)$ errors, where $\gamma_0$ is given by (\ref{eq:gast})
and
  \begin{equation}\label{eq:tf}
   \tilde F_{n,1}(\gamma)=
-\gamma n\log x+\log\biggl(1+2\sqrt{n\sum_{i=2}^n\binom ni
    x^{i+1}}\biggr)
  \end{equation}
where $x$ is the only positive root of the equation
  $$
   \frac { \sum_{i=2}^n(i+1)\binom ni x^{i+1}}
  {2n\sum_{i=2}^n\binom ni x^{i+1}+\sqrt{n\sum_{i=2}^n\binom ni x^{i+1}}}
   =\gamma .
  $$
\end{theorem}
\begin{IEEEproof} It is obtained by maximizing the function $g(\bfz)$ over the region
  $$
    \sM(1,\gamma)=\{\bfz\in \cZ_n: \sum_{i=1}^n iz_i=\gamma n, z_1=\sum_{i=2}^n
   z_i\}.
  $$
The Lagrangian takes the form
  $$
h(\bfz)+\sum_{i=2}^n z_i\Big(\log n+\log\binom ni\Big)+\lambda
\Big(\sum_{i=2}^n(i+1)z_i-\gamma n\Big),
  $$
where $\bfz=(z_1,z_2,\dots,z_n,1-\sum_i z_i)$ and
  $
  z_1=\sum_{i=2}^n z_i
  $
and $\lambda$ is an arbitrary multiplier.
Setting the partial derivatives to zero, we find the value $\lambda$ to satisfy
$2^x=\lambda,$ where $x$ is given above.
The calculations are tedious but straightforward and
will be omitted.
\end{IEEEproof}

The last theorem enables us to find the proportion of correctable errors
for the case when $A$ is the Hamming code of length $n=2^r-1, t=1. $
Since the examples below rely on the value of the ensemble-average distance,
we quote the corresponding result from \cite{bar08b}.
\begin{theorem}\label{thm:distance} \cite[Thm.5]{bar08b}
Let $\delta(\sH)$ be the asymptotic average relative distance of codes
in the $l$-hypergraph ensemble constructed from the local code $A$ of
length $n$ and distance $d_0.$ Then
   \begin{multline*}
   \delta(\sH)\ge \sup_{\omega>0}\Big\{\omega: \frac ln
      \log\frac{1+\sum_{i=d_0}^n\binom ni x_0^i}{x_0^{\omega n}}
  <(l-1)h(\omega)
\Big\}
  \end{multline*}
where
$x_0=x_0(\omega)$ is the positive solution of the equation
   $$
    \omega n+\sum_{i=d_0}^n\binom ni(\omega n-i)x^i=0.
   $$
\end{theorem}

For instance, for the case $n=31,l=5$ this theorem gives the value of
the relative distance $\delta(\sH)\ge 0.01618$ (the rate of codes
$R\ge 6/31$).
Performing the calculation in (\ref{eq:tf}), we find that the average
code from the ensemble $\sH(A, 5,m)$ the proportion of
errors correctable by codes in the ensemble using Algorithm II
to be at least $\gamma_0=1.2\times 10^{-5}.$

We include some more examples. In the following table $n=2^9-1.$

\vspace*{1mm}{\em Example 2:}

\vspace*{2mm}{\small\begin{tabular}{|c|c|c|c|c|}
\hline $l$  & 17 & 23 & 28 & 34 \\
\hline
Rate  & $0.7006$ & $0.5949$ &
$0.5069$ & $0.4012$  \\
\hline
$\gamma_0$ &
$0.000235$ & $0.000401$  & $0.000521$ & $0.000644$  \\
\hline $\delta(\sH)$  &0.00415  &0.00504  &0.00558 &0.00608  \\
\hline
\end{tabular}\\[1mm]
\hspace*{3.5mm}\begin{tabular}{|c|c|c|c|}
\hline
$l$  & 40 & 45 & 51\\\hline
Rate  & $0.2955$ & $0.2074$ & $0.1018$\\\hline
$\gamma_0$ & $0.000747$  & $0.000821$  &  $0.000898$ \\
\hline $\delta(\sH)$  &0.00648  &0.00676  &0.00704 \\
\hline
 \end{tabular}}

\vspace*{2mm}
It is also of interest to compute the values of $\gamma_0$ for code
rate $R(C)\approx 0.5.$

\vspace*{2mm}{\small
\begin{tabular}{|c|c|c|c|c|}
\hline
$n$ & $127$ &  $255$ & $511$ & $1023$ \\
  \hline
  $l$  & $9$ & $16$ & $28$ & $51$ \\
  \hline
  Rate  & $0.5039$ & $0.4980$ & $0.5068$ & $0.5015$ \\\hline
  $\gamma_0$ & $0.0002012$ & $0.0004873$  & $0.0005207$ & $0.0004227$ \\
\hline $\delta(\sH)$  &0.01157  &0.008658  &0.005581 & 0.003394  \\\hline

\end{tabular}}

\vspace*{2mm} These estimates are at least an order of magnitude
better than the corresponding results
in \cite{bur08,zya09}
obtained for LDPC codes and their generalizations based on
the ``flipping'' algorithm of \cite{zya75}.

\vspace*{.1in}
\emph{The case of large $n$.} As in the previous section, it is interesting to examine the case of long local
codes $A$ because it reveals some parallels with the analysis of the decoding
algorithm in the case of nonrandom hypergraphs \cite{bar08b}.
We begin with the observation that the proportion $\gamma_0$ of
correctable errors for the ensemble $\sH(A,t,m)$
computed above is a function of the number of errors $t$ that each local
code corrects in each iteration.
\begin{lemma}\label{lemma:hl}
Let $t=\tau n$, $d_0=\delta_0 n.$ The ensemble $\sH(A,t,m)$ contains codes that correct $\gamma N$ errors
for any $\gamma < \gamma_0(\tau) \triangleq \min(\tau,x_0(\tau))$
where
\begin{align*}
x_0(\tau) =& \sup\{x>0:\Big(1-\frac x{\delta_0}\Big)
    h\Big(\frac{x\tau}{\delta_0-x}\Big)
        +\frac{x}{\delta_0}h(\delta_0-\tau)\\
   &+\varepsilon_n <(1-1/l)h(x)\}
\end{align*}
and $\varepsilon_n=\log n /n.$
\end{lemma}
\begin{IEEEproof}
Referring to the proof of Lemma \ref{prop:hg}, we aim at
establishing conditions for the exponent $L$ of the event $\sL(\cE)$
to be negative as $m$ approaches infinity.
We assume that $\gamma\le \tau$ (otherwise
our estimates do not imply that
the convergence condition of Lemma \ref{lemma:red} holds with high
probability over the graph ensemble).

From (\ref{eq:L}), (\ref{eq:g}) we have
    $$
    L\le -(l-1)h(\gamma)+l\max_{\bfz\in\sM(t,\gamma)}
           \sum_{i=0}^nz_ih\Big(\frac in\Big)+\frac{l\log n}{n},
    $$
where $\sM(t,\gamma)$ is defined in (\ref{eq:mts}).
Next, write
  \begin{align}
   \sum_{i=0}^t z_ih\Big(\frac in\Big)\le \lambda
 h\Big(\frac {\sum_{i=1}^tiz_i}{\lambda n}\Big) =\lambda h
\Big(\frac {\mu_1}{\lambda}\Big), \label{eq:conv}
   \end{align}
where we have denoted $\sum_{i=0}^t z_i=\lambda,$ $\sum_{i=1}^t iz_i=\mu_1 n.$
In addition let us put $\sum_{i=d_0-t}^n iz_i=\mu_2 n,$ then
the values of the sums $\sum_i z_i$ and $\sum_i iz_i$ over each of the
three intervals $I_1=[0,t],\,I_2=[t+1, d_0-t-1],\, I_3=[d_0-t,n]$ can be found
from the following table:

\vspace{.1in}\begin{tabular}{cccc}
&$I_1$&$I_2$&$I_3$\\[1mm]
$\sum z_i$& $\lambda$& $1-\lambda-{\mu_1}/{\tau}$& $\mu_1/\tau$\\[1mm]
$\sum \frac i n z_i$&$\mu_1$&$\gamma-\mu_1-\mu_2$&$\mu_2$.
\end{tabular}

\vspace{.1in}
\noindent The variables introduced above depend on the point $\bfz$
and satisfy the
following natural constraints: for any $\bfz\in\sM(t,\gamma)$,
   \begin{align}
     \mu_1&\le\tau\lambda\nonumber\\
\tau\Big(1-\lambda-\frac{\mu_1}{\tau}\Big)\le\gamma-&\mu_1-\mu_2\le
    (\delta_0-\tau)\Big(1-\lambda-\frac{\mu_1}{\tau}\Big)\nonumber\\
  (\delta_0-\tau)\frac{\mu_1}{\tau}\le&\mu_2 \le \frac{\mu_1}{\tau}.\label{eq:con}
  \end{align}
Proceeding as in (\ref{eq:conv}), we can estimate the sum on $z_i$ in
$L$ as follows:
   \begin{equation}
   \label{eq:u1}
  \sum_{i=0}^nz_ih\Big(\frac in\Big)\le
 f(\lambda,\mu_1,\mu_2)
   \end{equation}
where
   \begin{align*}
   f(\lambda,\mu_1,\mu_2)=&
 \lambda h\Big(\frac{\mu_1}{\lambda}\Big)+
\Big(1-\lambda-\frac{\mu_1}{\tau}\Big)h\Big(\frac{\gamma-\mu_1-\mu_2}
  {1-\lambda-(\mu_1/\tau)}\Big)\nonumber \\&
  +\frac{\mu_1}\tau h\Big(\frac{\mu_2\tau}{\mu_1}\Big).
    \end{align*}
Our plan is to prove that some of the inequalities in (\ref{eq:con})
can be replaced by equalities, thereby
expressing the variables $\lambda,\mu_1,\mu_2$ as functions of $\gamma,\tau.$
We will rely on the fact that the function $f$ is concave in its domain,
proved in the end of this section.

Note that for all $\bfz\in \cZ_n$ the sum
  $$
   \sum_{i=0}^nz_ih\Big(\frac in\Big)\le h(\gamma)
  $$
and that it equals $h(\gamma)$ at the point
$\tilde\bfz$ such that $z_i=1$ for $i=\lceil\gamma n\rceil$ and $z_i=0$ elsewhere.
Also note that
since $\gamma<\tau,$ the point $\tilde\bfz$ is outside the region
$\sM(t,\gamma)$ and thus, by concavity,
  $$
    a:=\max_{\bfz\in\sM(t,\gamma)}
           \sum_{i=0}^nz_ih\Big(\frac in\Big)< h(\gamma).
  $$
Let $\bfz_1$ be the point at which this maximum is attained, and let
$\bfx_1=(\lambda,\mu_1,\mu_2)$ be the corresponding point for the
arguments of $f.$ By construction, the point $\bfx_1$ satisfies the
inequalities of (\ref{eq:con}).
At the same time, consider the function $f(\cdot)$ on the line
$\lambda=\mu_1=\mu_2.$ As the variables approach 0 along this line, the
value $f(\lambda,\mu_1,\mu_2)$ approaches $h(\gamma).$

To summarize, we have found two points, $\bfx_1$ and
$\bfx_2=(0,0,0)$ that are located on different sides of the hyperplane
   $$
    \tau\Big(1-\lambda-\frac{\mu_1}{\tau}\Big)=\gamma-\mu_1-\mu_2
   $$
such that $f(\bfx_1)\ge a, f(\bfx_2)>a.$ Invoking concavity of the function
$f,$ we now conclude that there is a feasible point $\bfx'$ on this hyperplane
such that $f(\bfx')\ge a.$

Therefore, put $\mu_2=\gamma-\tau(1-\lambda)$ and write
  \begin{align*}
    f_1(\lambda,\mu_1)=&\lambda h\Big(\frac{\mu_1}{\lambda}\Big)+
  \Big(1-\lambda-\frac{\mu_1}{\tau}\Big)h(\tau)\\
   &+\frac{\mu_1}\tau h\Big(\frac{\tau(\gamma-\tau(1-\lambda))}{\mu_1}\Big)
  \end{align*}
where the variables are constrained as follows: for any $\bfz\in\sM(t,\gamma),$
  \begin{align}
     \mu_1&\le\tau\lambda\nonumber\\
    \tau(1-\lambda)&-\mu_1\ge 0\label{eq:e1}\\
  (\delta_0-\tau)\frac{\mu_1}{\tau}\le\gamma-&\tau(1-\lambda)
                 \le \frac{\mu_1}{\tau}.         \label{eq:e2}
  \end{align}
Since $f_1$ is a restriction of $f$ to a hyperplane, it is still
concave.
Now notice that $f_1(1,\tau)=h(\gamma)$ and that the point $(1,\tau)$
does not satisfy inequality (\ref{eq:e1}) and the left of the inequalities
(\ref{eq:e2}). Repeating the above argument, we claim that
the function $f$ in (\ref{eq:u1}) can be further restricted to the intersection of the
planes $\tau(1-\lambda)=\mu_1$ and $(\delta_0-\tau)(\mu_1/\tau)=
\gamma-\tau(1-\lambda).$ Altogether this gives:
  $$
  \lambda=1-\gamma/\delta_0, \;\;\mu_1=\gamma\tau/\delta_0.
  $$
Let us substitute these values into the expression for $f_1$ and
rewrite (\ref{eq:u1}) as follows: for any $0\le\gamma<\tau,$
  \begin{equation}\label{eq:2}
    \max_{\bfz\in\sM(t,\gamma)}\sum_{i=0}^nz_ih\Big(\frac in\Big)\le
     \Big(1-\frac{\gamma}{\delta_0}\Big)h\Big(\frac{\gamma\tau}{\delta_0-\gamma}
    \Big)+\frac\gamma{\delta_0}h(\delta_0-\tau).
  \end{equation}
Thus if the condition in the statement is fulfilled then $L<0$.
This concludes the proof.
\end{IEEEproof}

\vspace*{.1in}
{\em Remark.} The main part of the proof is estimating the solution
of the following linear program
  \begin{align*}
   \max_{\bfz} &\sum_{i=1}^n z_i h\Big(\frac in\Big)\\
   \bfz=(z_0,z_1,&\dots,z_n)\in \sM(t,\gamma)
  \end{align*}
where the variables define a probability distribution
on $\{0,1,\dots,n\}.$
It is clear from concavity that the maximum is attained at the point
where among all the indices $i\in I_1$ at most one value $z_i$ is nonzero,
and the same applies to $I_2$ and $I_3.$ We have shown that the value of the
program is bounded above by the right-hand side of (\ref{eq:2}). The following
point gives this value and is therefore a maximizing point:
  $$
   z_{i_1}=1-\frac\gamma{\delta_0},\; z_{i_2}=\frac\gamma{\delta_0}, \;
  z_i=0 \text{ otherwise},
  $$
where $i_1=n\gamma\tau/(\delta_0-\gamma), i_2=n(\delta_0-\tau).$
Since
  $$
   \frac{\gamma\tau}{\delta_0-\gamma}\le \tau,
  $$
this shows that the worst-case allocation of errors to vertices in a given
part of the graph assigns no edges to vertices that are neither good
nor bad. This also confirms the intuition suggested by Lemma \ref{lemma:red}
that bad vertices (vertices assumed to add errors) should each be assigned
the smallest possible number of error edges $d_0-t.$

The next proposition is now immediate.
\begin{proposition}\label{prop:th}
The ensemble $\sH(A,l,m)$ with long local codes
contains codes that can be decoded using
Algorithm II to correct all error patterns whose weight is less than
$\gamma_0 N,$ where
  \begin{equation}\label{eq:gt}
   \gamma_0=\max_{0<\tau\le\delta_0/2} \gamma_0(\tau).
  \end{equation}
\end{proposition}

Estimating the number of correctable errors for the ensemble $\sH(A,l,m)$
from Proposition \ref{prop:th}
analytically is difficult because it involves
optimization on $\tau$ (generally, the local codes
should be used to correct a smaller than $\delta_0/2$ proportion of errors).
We note that in the particular case of $\tau=\delta_0/2$ the proof of
Lemma \ref{lemma:hl} can be considerably simplified, although the resulting
value of $\gamma$ is not always optimal.

{\em Example 3.} Let  $l=3.$
Using local codes with $\delta_0=0.05$ we can construct hypergraph codes
of rate $R\ge 0.19$. From \cite[Cor.~6]{bar08b}, the ensemble-average
relative distance is at least $\delta\approx0.0112$ and the proportion of errors
correctable by Algorithm II is found from (\ref{eq:gt}) to be
$\gamma_0\approx 0.0035.$

{\em Example 4.} Let $\delta_0=0.01$ and $l=10.$
In this case, we find from \cite[Cor.~6]{bar08b} the value of the
relative distance
$\delta\approx 0.00599.$
The code rate satisfies $R\ge 0.14.$
Performing the computations in (\ref{eq:gt}) and Lemma \ref{lemma:hl}
we find the estimate of the proportion of correctable errors to be
$\gamma_0\approx 0.002198$.

\emph{Proof that $f(\lambda,\mu_1,\mu_2)$ is concave.} First we prove
that the function
  $$
    \phi(x,y)=(1-x)h\Big(\frac{\gamma-y}{1-x}\Big)
  $$
is concave (not necessarily in the strict sense) for $0<x,y<1,0<\gamma-y<1-x.$
For that, let us compute its Hessian matrix:
  $$
   H=\frac1{\ln 2}\begin{pmatrix}\frac{\gamma-y}{(1-x)(\gamma-y+x-1)}&
- \frac1{\gamma-y+x-1}\\[2mm]
      - \frac1{\gamma-y+x-1}&\frac{1-x}{(\gamma-y)(\gamma-y+x-1)}
   \end{pmatrix}
  $$
The eigenvalues of $H$ are
  $$0,\;\;\frac{(\gamma-y)^2+(1-x)^2}{(1-x)(\gamma-y)(\gamma-y-(1-x))}<0,
  $$
so
$H\preceq 0,$ and so $\phi$ is concave.
Next observe that the function
  $$
    \Big(1-\lambda-\frac{\mu_1}{\tau}\Big)h\Big(\frac{\gamma-\mu_1-\mu_2}
  {1-\lambda-(\mu_1/\tau)}\Big)
  $$
can be obtained from $\phi$ by a linear change of variables
  $$x=\lambda+\mu_1/\tau, \;\;y=\mu_1+\mu_2$$
and therefore is also concave.
Finally, the functions $\lambda h(\mu_1/\lambda)$ and
$(\mu_1/\tau)h(\mu_2\tau/\mu_1)$ are also concave, and thus so
is the function $f(\lambda,\mu_1,\mu_2).$

\section{Conclusion}
We have estimated the proportion of errors correctable by codes from
ensembles defined by random $l$-partite graphs, $l\ge 2.$
In contrast to the case of expander codes \cite{sip96},
\cite{zem01}, \cite{bar05},  \cite{bil04}, \cite{bar08b}
our calculations cover the case of local codes of arbitrary given length
and distance, including small values of the distance.
The behavior of code ensembles considered here was examined from a
different perspective in \cite{bar08b} where we computed estimates
of the expected distance and weight distribution of these codes.
The paper \cite{bar08b} and the present work together
provide answers to the set of basic questions regarding random networks of
short linear binary codes and extend our perspective of concatenated
code constructions to the case of sparse regular graphs.


\vspace*{2mm}{\em Acknowledgment.} The authors are grateful to J{\o}rn Justesen
and to an anonymous reviewer for useful comments on this work.

\providecommand{\bysame}{\leavevmode\hbox to3em{\hrulefill}\thinspace}
\providecommand{\href}[2]{#2}


\end{document}